# Comparison of Full-text Versus Metadata Searching in an Institutional Repository: Case Study of the UNT Scholarly Works

| Laura Waugh | Hannah Tarver | Mark Phillips | Daniel Alemneh |
| --- | --- | --- | --- |
| University of North Texas | University of North Texas | University of North Texas | University of North Texas |
| 1155 Union Circle #3015190 | 1155 Union Circle #3015190 | 1155 Union Circle #3015190 | 1155 Union Circle #3015190 |
| Denton, TX 76203-5017 | Denton, TX 76203-5017 | Denton, TX 76203-5017 | Denton, TX 76203-5017 |
| laura.waugh@unt.edu | hannah.tarver@unt.edu | mark.phillips@unt.edu | daniel.alemneh@unt.edu |

**ABSTRACT**

Authors in the library science field disagree about the importance of using costly resources to create local metadata records, particularly for scholarly materials that have full-text search alternatives. At the University of North Texas (UNT) Libraries, we decided to test this concept by answering the question: What percentage of search terms retrieved results based on full-text versus metadata values for items in the UNT Scholarly Works institutional repository? The analysis matched search query logs to indexes of the metadata records and full text of the items in the collection. Results show the distribution of item discoveries that were based on metadata exclusively, on full text exclusively, and on the combination of both. This paper describes in detail the methods and findings of this study.

**Keywords**

*Metadata, full-text, information retrieval, institutional repositories*

## 1. Background

Metadata creation in digital libraries can be costly, time consuming, and may require technical expertise. For these reasons, among others, authors in the library science field often disagree about the importance of creating local metadata records. Opinions vary, including proponents of metadata creation (e.g., Albitz, 2014; Beall, 2008; Kostoff, 2010) -- citing opportunities for variant words, phrases, or spellings, translations of foreign texts, and increased search precision -- and those who argue that metadata can be replaced by full-text searching and other technologies for time and cost efficiency in digital preservation and increased possible matches in a search (e.g., Albitz et al., 2014; Rosenthal, 2014). At the University of North Texas (UNT) Libraries, we decided to test this concept by answering the question: What percentage of search terms retrieved results based on full-text versus metadata values for items in the UNT Scholarly Works institutional repository (IR)? We chose the UNT Scholarly Works collection because these items are scholarly in nature, are primarily text-based, and most are modern or born-digital documents that will have easily-searched OCR files.

UNT Scholarly Works is a designated collection within the larger UNT Digital Library, meant to collocate materials that directly reflect the research, creative, and scholarly activities of UNT community members. Current faculty and staff members can submit items that represent the scholarship of their field; most commonly items include journal articles, presentation slides, book chapters, conference posters, or reports. The UNT Libraries have chosen not to utilize self-submission software to assist with items for the IR. Materials are emailed directly to a repository librarian and trained staff members create metadata records in accordance with established input guidelines (UNT Libraries, 2015) to generate a locally-qualified Dublin Core record for each item. All metadata records meet various guidelines and standards, including name authority control – managed by the UNT Name App[1] for creator, contributor, and publisher names – the Library of Congress Extended Date/Time Format (EDTF) (Library of Congress, 2015) for dates, and controlled vocabularies (such as Library of Congress Subject Headings) for subject terms when appropriate. Supplied keywords or subject terms from the author(s) or publisher are also included in the metadata when available. These standards ensure some level of quality control and consistency for records in the UNT Scholarly Works collection.

This study is entirely quantitative, focused on answering *how* users reach items for which they view the metadata or full text of the scholarly materials. For each of the cases analyzed in this paper, a user clicked on an item link from a search results list, which displays title, author(s), date, and content description for each matching item. This suggests some level of relevance for the user, however the scope of this research does not attempt to answer questions aside from query matches; in particular it does not address metadata quality (e.g., whether inappropriate subjects or inaccurate information in metadata records could affect the search results) or user satisfaction (e.g., if users felt the results were reasonable or expected).

---

[1] UNT Name App. http://digital2.library.unt.edu/name/

## 2. Method

This analysis used Web server logs from the application server that provides access to the UNT Digital Library. The log files were limited to discoveries of items in the UNT Scholarly Works collection -- 3,651 unique items at the time this research was conducted -- which occurred between May 4, 2014 and January 24, 2015. The raw dataset contained 172,115,682 lines during that timeframe, in the standard Extended Log File Format. Further limitations removed requests made by known robots, or without known search queries, resulting in a two-column intermediary dataset that contained 3,797 item-query pairs (i.e., a local identifier for each discovered item and the request used). Following normalization, the final dataset contained 2,343 *unique* item-query pairs.

The UNT Digital Library uses the Solr full-text indexer, which can provide explanatory information noting why a query yielded certain results, if a specific document would be returned by a query, and in which specified fields the terms appear. To utilize this, the item-query pairs were fed to the Solr search system. The final dataset used in the remainder of this paper lists the percentage of each search query that was found in the metadata (full descriptive record), full text, and the four specified fields -- title, subject, agent (both creator and contributor values) and description (see Table 1). The dataset has 2,341 query results; two incomplete samples were discarded during processing.

**Table 1.** Example dataset entries for three search queries.

| Dataset Field | Example 1 | Example 2 | Example 3 |
|---|---|---|---|
| Item | metadc129697 | metadc146510 | metadc155618 |
| Query | susan cheal | human trafficking | article writing |
| Query Tokens | 2 | 2 | 2 |
| Metadata | 100% | 100% | 50% |
| PageText | 0% | 100% | 100% |
| Title | 0% | 100% | 50% |
| Subject | 0% | 100% | 0% |
| Agent | 100% | 0% | 0% |
| Description | 0% | 100% | 0% |

## 3. Findings

First, some basic analysis revealed statistical facts about the data collected. For example, the dataset represented 1,448 unique items, comprising 39% of the Scholarly Works collection. Items in the dataset were queried an average of 1.62 times during the time period, however, the actual rate ranged from a single query to 15 queries for a specific item (see Table 2).

**Table 2.** Statistics for the number of queries per unique item.

| N | min | median | max | sum | mean | stddev |
|---|---|---|---|---|---|---|
| 1,448 | 1 | 1 | 15 | 2,341 | 1.62 | 1.28 |

Although queries varied in length, they were analyzed as individual words (or tokens) rather than phrases. This allowed for partial matches in a given field, resulting in percentages less than 100. The distribution of tokens across queries ranged from 1 to 20 tokens (see Table 3).

**Table 3.** Statistics for the number of tokens per query.

| N | min | median | max | sum | mean | stddev |
|---|---|---|---|---|---|---|
| 2,341 | 1 | 3 | 20 | 4,928 | 2.11 | 1.48 |

At this point, the analysis turned toward answering the research question. Table 4 gives a breakdown of the total number of queries found in the metadata and full text including partial and full matches. The numbers overlap in cases where tokens appeared in both indexes. A number of record discoveries were dependent entirely on the full text (200) or metadata (382).

**Table 4.** Record discoveries based on matches in metadata and full text. (n=2341)

| Matches found in: | Total queries found: |
|---|---|
| Both metadata and full text | 1,759 |
| Any part of query in full text | 1,959 |
| 100% of query in full text | 1,830 |
| Queries ONLY in full text | 200 |
| Any part of query in metadata | 2,141 |
| 100% of query in metadata | 1,888 |
| Queries ONLY in metadata | 382 |

For the subset of record discoveries with token matches in both indexes (1,759), Table 5 shows how many items could be found using either index equally, how many had a partial match in one index, with a full match in the other index, and the number of queries (21) that could be found *only* through the combination of metadata and full-text versus either index alone.

Table 5. Record discoveries categorized for discoveries in both metadata and full text. (n=1759)

| Overlapping matches found in: | Number of queries: |
|---|---|
| Metadata/ full text equally (m=100/p=100) | 1,403 |
| ONLY with metadata/ full text combined (m<100/p<100) | 21 |
| Metadata/ partial full text (m=100/p<100) | 106 |
| Partial metadata/ full text (m<100/p=100) | 229 |

At a more granular level, Table 6 displays average query matches by field: title, subject/keyword (Subj.), agent, or content description (Descr.). Most terms were found in the agent and title fields, suggesting that many of the queries were looking for specific items rather than for items with a general topic.

Table 6. Average percentage of query found for each record discovery. (n=2341)

| | General Location | | Individual Metadata Fields | | | |
|---|---|---|---|---|---|---|
| Location of Query Match | Page Text | Metadata | Title | Subj. | Agent | Descr. |
| Average % of each query by field | 81.17 | 86.14 | 32.31 | 19.20 | 45.27 | 30.79 |

Table 7 shows record discoveries broken down by match percentages of each field, for the entire dataset. This shows the extent of the matches (partially for longer query strings) and the overlap across multiple fields.

Table 7. Record discoveries per field based on percentage of query present in field. (n=2341)

| | 0% | 1-49% | 50-74% | 75-99% | 100% | %>=1% found in field |
|---|---|---|---|---|---|---|
| Title | 1,438 | 86 | 181 | 26 | 610 | 38.57% |
| Subj. | 1,622 | 215 | 249 | 33 | 222 | 30.71% |
| Agent | 1,161 | 103 | 100 | 4 | 973 | 40.41% |
| Descr. | 1,434 | 115 | 223 | 36 | 533 | 38.74% |

Table 8 shows the match percentage by field for the subset of 382 items discovered through metadata-only retrieval. Most query matches were in the agent fields, followed by title and description.

Table 8. Number of record discoveries per field from queries only found in metadata. (n=382)

| | 0% | >=1% | %>=1% found in field |
|---|---|---|---|
| Title | 282 | 100 | 26% |
| Subj. | 318 | 64 | 17% |
| Agent | 156 | 226 | 59% |
| Descr. | 263 | 119 | 31% |

## 4. Conclusion

The findings of this work present a few interesting pieces of data. For the 2,341 item discoveries used as the dataset for this paper, most (75%) were based on queries matching both metadata and full text. However, 382 (16%) were entirely dependent on metadata while 200 (9%) were dependent entirely on full-text searching. This leaves 1,759 (75%) of the record discoveries being completed with text that was present both in the metadata and full text. A small number of items (21) could not have been retrieved without the combination of metadata and full-text searching. Other collections would likely yield different results. For example, a photograph collection will have no full-text index and relies solely on metadata; likewise, a digitized newspaper collection would most likely have less robust metadata and rely heavily on full-text searching.

Additionally, most of the dataset queries matched terms in agent (DC_Creator and DC_Contributor) or title fields, which leads us to believe that there are a number of "known item" lookups, for a specific title or author. The description field satisfies the third most queries. This is unexpected since the description is more likely to contain topical information, however the subject field had the fewest query matches suggesting less focus on general topics. It should be noted that the UNT Scholarly Works repository generally does not add subject headings to articles from standard controlled vocabularies such as LCSH but relies on keywords submitted with the paper or extracted from the text by the repository librarian.

Although many search queries had overlapping results in both the metadata and full-text, a number of item discoveries occurred only through metadata values or the combination of metadata and full-text. This suggests that creating local records does support item discovery and retrieval. Further research may offer additional information about where best to concentrate efforts, or the role of other value-added services that make use of metadata (e.g., faceted browsing) and that would not be possible with full-text searching alone.